\documentstyle[aps,prl]{revtex}
\begin{document}
\title{ Alternating Spin and Orbital Dimerization in
Strong-coupling Two-band Models}
\author{Swapan K. Pati and Rajiv R. P. Singh}
\address{Department of Physics, University of California, 
Davis, California 95616}
\author{Daniel I. Khomskii }
\address{Laboratory of Solid State Physics, University of Groningen,
Nijenborgh 4, 9747 AG Groningen}

\twocolumn[\hsize\textwidth\columnwidth\hsize\csname
@twocolumnfalse\endcsname
\date{\today}
\maketitle
\widetext
\begin{abstract}
\begin{center}
    \parbox{6in} {We study a one-dimensional Hamiltonian 
consisting of coupled SU(2) spin and orbital degrees of freedom.
Using the density matrix renormalization group, we calculate
the phase-diagram and the ground state correlation functions
for this model. We find that, in addition to the 
ferromagnetic and power-law antiferromagnetic phases
for spin and orbital degrees of freedom,
this model has a gapless line
extending from the ferromagnetic phase to the Bethe ansatz
solvable SU(4) critical point,
and a gapped phase with doubly degenerate ground states which form
alternating spin and orbital singlets.
The spin-gap
and the order parameters are evaluated and the relevance to
several recently discovered spin-gap materials is discussed.}

\end{center}
\end{abstract}

]

\narrowtext

In recent years a remarkable number of new materials have been synthesized,
which despite having magnetic ions have non-magnetic ground states,
a finite correlation length and a gap in the spin-excitation spectrum.
Among them, the spin-ladder forming Strontium Cuprates\cite{srcuo}, the
spin-Peierls systems Copper Germanates\cite{cugeo3} and the periodically
depleted Calcium Vanadates\cite{cav4o9} have attracted the greatest attention.
Here we discuss a novel spin-gap phase, which although
mathematically closely related to the spin-ladder and
spin-Peierls systems, is physically quite different. It arises
in systems with spin and orbital degeneracy and requires coherence
over neighboring atoms of both degrees of freedom. We also motivate
such a spin-orbital model from a quantum-chemical analysis of
recently discovered spin-gap 
materials Na$_2$Ti$_2$Sb$_2$O\cite{kauzlarich} and Na$_2$V$_2$O$_5$
\cite{nav2o5} and discuss the relevance of this new phase to them.

The two-band Hubbard model is well-known in context of
magnetic insulators with Jahn-Teller ions \cite{kugel-khomskii}.
The Hamiltonian is:
\begin{eqnarray}
H=\sum
t_{ij}^{\alpha\beta}c_{i\alpha\sigma}^+c_{j\beta\sigma}+
\sum_{(\alpha\sigma)\neq(\beta\sigma')}
U_{\alpha\beta}n_{i\alpha\sigma}n_{i\beta\sigma'} ,
\end{eqnarray}
where $i,j$ are site indices, $\alpha,\beta=1,2$ the orbitals
and $\sigma,\sigma^\prime$ the spin indices. 
Quarter filling of the bands amounts to one electron per atom.
In the strong coupling limit this system is a Mott insulator,
and the state of each ion can be characterized by a spin $S_i$
and the orbital state can be mapped into a pseudospin $T=1/2$
so that orbital one corresponds to $T^z=1/2$ and orbital 2
to $T^z=-1/2$. Thus in the strong coupling limit,
the effective spin-pseudospin Hamiltonian in one-dimension
becomes \cite{kugel-khomskii}:
\begin{eqnarray}
{\cal H}= &J_1\sum_i \vec S_i\cdot\vec S_{i+1}
          +J_2\sum_i (\vec T_i\cdot\vec T_{i+1}+A T_i^zT_{i+1}^z)
\nonumber \\
          &+K\sum_i \vec S_i\cdot\vec S_{i+1} (\vec T_i\cdot\vec T_{i+1}
          +B T_i^z T_{i+1}^z)
\end{eqnarray}
The simplest assumptions $t^{11}=t^{22}=t,t^{12}=0$, together with
a single $U$ leads to
\begin{eqnarray}
H= \sum_i \big(J_1 \vec S_i\cdot\vec S_{i+1}+J_2 \vec T_i\cdot \vec T_{i+1}
\nonumber \\
+K(\vec S_i\cdot \vec S_{i+1})(\vec T_i\cdot \vec T_{i+1})\big)
\end{eqnarray}
with $J_1=J_2=K/4$. This special point has SU(4) symmetry.
However, there are many ways in which one can
find deviations from these parameters.
For example, it was shown by Arovas and Auerbach\cite{arovas}
in the context of $C_{60}-TDAE$, that more than one $U_{\alpha\beta}$ leads to 
the Hamiltonian in Eq.~3 with the 
$SU(2)\times SU(2)$ symmetric parameter
space reducing to $J_1+J_2=K/2$, but with $J_1$ not necessarily equal to $J_2$.

Here we study the ground states of this model with $SU(2)\times SU(2)$ symmetry
with arbitrary $J_1$, $J_2$ and positive $K$.
The calculated ground-state phase diagram is shown 
in Fig.1, where the Arovas-Auerbach line, $J_1+J_2=K/2$ 
is shown by the dotted line (which goes through point A).
It is well known that the presence of orbital degrees of freedom can alter the
nature of spin order, giving rise to both ferromagnetism and
antiferromagnetism \cite{kugel-khomskii}.
The ground states in phases I, II and III are known exactly.
They are direct products of a spin and a pseudospin ground states.
In phase I both spin and pseudospin
degrees of freedom are fully
polarized ferromagnets. In phase II the pseudospins are ferromagnetic
whereas the spins are antiferromagnetic and their ground state is the
Bethe Ansatz ground state of the spin-half antiferromagnetic
chain. In phase III, spins and pseudospins are
interchanged with respect to phase II. All these 3 phases are
conventional magnetic phases appropriate for 1D, with
gapless spin excitations. We will concentrate in the rest of the paper
on the region IV, which we show has an excitation-gap, 
and the critical line AB, where gapless excitations exist.
There are two other points in the phase diagram, where the ground state
is known exactly. The point A has SU(4) symmetry and is also Bethe-ansatz
integrable \cite{sutherland}. 
It has gapless excitations and power-law correlations.
The exact ground state for point C was recently obtained by
Kolezhuk and Mikeska \cite{kolezhuk}. This ground state is doubly degenerate and
there is a gap in the excitation spectrum.

To determine the properties of the region IV and the line AB, we turn
to numerical methods. We use the density matrix renormalization
group (DMRG) method\cite{white} to calculate the ground state, the 
excitation gap and the spin and orbital correlation functions.
The overall features of the method remain essentially the same for this problem 
as for the Heisenberg spin problems\cite{dmrg}. 
However, to treat the biquadratic term properly, and to take advantage
of the enhanced symmetry, modifications are needed.
It is known that the product of
operators within the same block in the block density matrix
eigenvector basis will be error prone, because the resolution of
identity in the density-matrix basis is non-trivial\cite{product}. 
This creates a difficulty in renormalizing the biquadratic term.
To circumvent
this difficulty, we have carried through any operator products required
to construct the Hamiltonian matrix at every iteration. Besides, as 
the system posseses a global $SU(2) \times SU(2)$ symmetry, we define 
a single site with 4-states from the spin and orbital
degrees of freedom and
ensure that at every iteration both $S_{tot}^z$ and $T_{tot}^z$ 
are preserved as good quantum numbers.

We have used periodic boundary conditions throughout this study
and have verified our results extensively. The ground state energy 
for point C (see Fig.1) is correct to numerical 
accuracy and the lowest excitation 
gap calculated to be $0.3756$ in units of $K$ compares with the variational
estimate of $0.375$\cite{kolezhuk}, with the DMRG cut-off, $m=100$. 
At the SU(4) point (point A in Fig.1), the ground state 
energy is accurate with that obtained
by the Bethe {\it ansatz}\cite{sutherland} to fourth significant decimal place
and the
excitation gap vanishes to numerical accuracy. 
We also verify the ground state energies and quantum numbers
of phases I, II and III and obtain their phase boundaries by DMRG.
In all our calculations described below, we have kept the cut-off
$m$ to be $120 < m <150$. All the calculations are repeatedly checked
by exact diagonalization results for small system sizes and we 
report here the results for system sizes upto $N \ge 30$. Note
that, each site contains a spin and an orbital variable.

We have calculated the excited state energies in the subspace
of $(S_z^{tot}, T_z^{tot})=(1,1)$ while the ground state remains
in $(S_z^{tot},T_z^{tot})=(0,0)$ subspace. For the line BA
as well as for the whole of the incommensurate line(line AD in 
Fig.1), 
we have calculated the excitation gaps from the $N=4n$ data. This is
to avoid many-fold ground-state degeneracy for $N=4n+2$ \cite{ueda}.
Furthermore, to obtain these gaps in the thermodynamic limit, 
the calculated finite-size gaps, for the $N=4n$ systems are fitted with 
a function of the form
\begin{equation}
\Delta(N)= \Delta +A/N + B/N^2 +C/N^3 + \ldots
\end{equation}

We find that the SU(4) critical point A is the endpoint of
a critical line AB, where the 
point B, is a special point where several ground
states, including the fully ferromagnetic spin and pseudospin
ground states become exactly degenerate.
At the SU(4) point, we have verified
that the spin structure factor peaks at $q=\pi/2$ and the decay of the
power-law
real-space correlations are consistent with the $3/2$-power \cite{affleck}.
Furthermore, all along the open interval AB, the
structure factor peak remains at $q=\pi/2$.
Also on this open interval, the ground state is a
singlet and the spin and pseudospin gap remains zero. 
Numerically we find that the gap opens with an exponent of
$1.5 \pm 0.25$ along the line AD close to the point A. 
 
The point C is known to have degenerate ground states which can be
written as a matrix-product consisting of alternating
singlet bonds in both the spin and orbital variables \cite{kolezhuk}.
There is a finite correlation length and the ground state spin structure 
factor peaks at $q=\pi$.
Along the line AC above the SU(4) point A,
the ground state remains doubly degenerate with a spin-gap (see Fig.1).
This gap goes through a maximum at $J=J_1=J_2=(0.5 \pm 0.02)K$, while 
going from point C to point A where it vanishes to zero.
Along this line, from point A
to the point where the gap becomes maximum (point D in Fig.1), the spin as 
well as the pseudospin pair
correlations are incommensurate. The peak in the structure factor moves 
from $q=\pi$ at D to $q=\pi/2$ at A. 

Consider now the line which runs from point C to infinity with
equal $J_1/K$ and $J_2/K$ values. All along this line,
the spin-correlations stay peaked at ($q=\pi$) and there is a finite
spin gap. In the small $K/J$ limit, the results are
similar to those found for the ordinary spin ladder, where a gap opens
for any finite interchain coupling. Our results are consistent with
those of Nersesyan and Tsvelik\cite{tsvelik} and Mostovoy\cite{mostovoy}
 in that the gap is linear in $K/J$. In Fig.2, we plot the lowest 
excitation gap in units of $J=J_1=J_2$, as a function of $K/J$.

Over the entire region IV, 
the ground state with $N=4n$
is doubly degenerate. It is a spin and pseudospin singlet,
with a finite excitation-gap. These results are 
in accordance with the Lieb-Schultz-Mattis theorem\cite{lsm}.
Furthermore, it has the same broken 
symmetry as the exactly solved point C. This is verified by
calculating the square of the order parameter
\begin{eqnarray}
 Q^2(i-j)= & < (S_i^z S_{i+1}^z -c)( S_j^z S_{j+1}^z -c) 
\nonumber \\
&+ (T_i^z T_{i+1}^z 
-c^{\prime})(T_j^z T_{j+1}^z-c^{\prime}) >
\end{eqnarray}
for large $i-j$, where $S_i^z$ and $T_i^z$ are the $z$-component of
spin and pseudospin operators respectively. $c$ and $c^{\prime}$
are the average of two-particle 
correlations, i.e., $c=1/N \sum_i <S_i^z S_{i+1}^z>$
and $c^{\prime}=1/N \sum_i <T_i^z T_{i+1}^z>$, N is the number of
spin-orbital pairs. We find that $Q^2$ remains finite in this
phase. In the context of spin degrees of freedom, it is appropriate 
to mention here that the critical line (BA) and gapful phases of this model 
discussed above are similar to that of the Majumdar-Ghosh model for 
spin-1/2 chain with nearest and second neighbour exchanges\cite{mgmodel}.

We now turn to some recently synthesized spin-gap materials
for which this model is relevant.
The material Na$_2$Ti$_2$Sb$_2$O was recently found to have a
finite temperature phase transition at
$T_c\approx 110K$ \cite{kauzlarich}. Below this temperature the uniform
susceptibility drops sharply without any formation of
magnetic order, and shows activated (spin-gap) behavior.
In this material, the spin-half $Ti^{3+} (d^1)$ ions form a square
lattice. Thus, this behavior is in marked contrast to the undoped Cuprates,
which show antiferromagnetic order. 

This material has inverse ``K$_2$NiF$_4$" structure, consisting
of layers of oxygen ions forming square-lattice,
and Ti ions sitting in between them. The Sb
ions are located above and below the centres of oxygen plaquettes as
shown in Fig.3a. Each Ti ion is surrounded by four Sb ions and 
two oxygens, forming approximate octahedron. In this local
tetragonal coordination, the triply degenerate $t_{2g}$ levels
are split into a doublet $e_g$ and an $a_1$ levels. As the covalency
of the Ti$-$Sb bond gives the main contribution to the crystal-field
splitting, the $e_g$ doublets should lie lower. Thus the ground state
of Ti contains one d-electron in a doubly degenerate orbital, which
choosing the O-Ti-O axis as the z-axis are the $d_{xz}$ and the $d_{yz}$
orbitals. 
These orbitals overlap through strong $\pi$-hybridization with the
corresponding $p_\sigma$ orbitals of Sb thus giving rise to
two independent one dimensional structures extending along two
mutually perpendicular axes of the crystal (see fig.3b).
Thus to a first approximation this represents a quasi-1D
quarter-filled 2-band system.

Magnetic properties of the material Na$_2$Ti$_2$Sb$_2$O, 
are qualitatively consistent with the
gapped phase discussed here, although, more experiments are
clearly needed to investigate the relevant atomic orbitals \cite{pickett}
and the nature of the low temperature phase of this system. It is also
interesting to note that the 
isostructural material Na$_2$Ti$_2$As$_2$O (where As replaces Sb) in many
respect shows similar behavior above a certain temperature
with a gradual reduction in 
magnetic susceptibility as the temperaure is lowered but also shows
some evidence for ferromagnetism at very low temperature\cite{private}. This
could be expected from changing parameters in the spin-pseudospin
models discussed above.

Another material that has recently gotten considerable attention is
Na$_2$V$_2$O$_5$. This material is also a two-band quarter-filled
quasi-1D Hubbard system\cite{smolinski}. In this case, the two orbitals 
arise from two chains
of a ladder-like structure. Hence pseudospin ordering corresponds
to charge ordering on different atoms\cite{fulde,seo,mosto-khom}. 
It has been argued that in this material the spin-gap
may be entirely due to charge ordering\cite{mosto-khom}. We note that 
at the SU(4) point, the spin and pseudospin
correlations are peaked at $\pi/2$ and this is different 
from the period 1 and period 2 ( ferro and antiferromagnetic) charge
ordering scenarios, which have been discussed before.
In the spin-gap phase discussed
here, all atoms remain in an equivalent charge state. 
The question of whether such an alternating spin and orbital coherence
between neighbors plays any role in this material deserves further attention.

In conclusion, we have used the density matrix renormalization group
method to study the ground state, excitation-gap and
correlation functions of an $SU(2)\times SU(2)$ spin-orbital model.
We show that this model has a rich phase diagram.
In addition to conventional ferromagnetic and antiferromagnetic
phases in the spin and pseudospin variables,
there is a gapless
critical line which runs from the fully polarized ferromagnetic
phase to the Bethe-ansatz solvable SU(4) critical point.
We also find that over a wide parameter range this model
has a broken symmetry gapped phase, where the system forms
an alternating pattern of spin and orbital singlets. 
The relevance of this phase to various recently discovered spin-gap
materials is discussed.

Acknowledgements: We thank M. Mostovoy, S. Ramasesha and D. Sen
for valuable discussions. We have also benefitted from
earlier DMRG programs developed in S. Ramasesha's group. This work was
financially supported by a grant from NSF number DMR-96-16574.

{\bf Figure Captions:}

\vspace*{0.5cm}

Fig.1: Phase Diagram for the model in Hamiltonian~(3) in the $J_1/K$, 
$J_2/K$ parameter space. Thick solid lines are 1st-order phase boundaries
and BA is a critical line. See text for details. \\

Fig.2: Lowest excitation gap in units of $J$ as a function of $K/J$ for 
the line from point A to infinity through point D and C as in Fig.1. \\

Fig.3a: Structure of the material Na$_2$Ti$_2$Sb$_2$O. \\

Fig.3b: Cross-sectional view of the Titanium $d_{xz}$ 
and $d_{yz}$ orbitals, together with the p$_z$-orbitals of Sb in the X-Y plane.
As shown, hoppings $t^{11}$ (shaded-shaded)$=$ $t^{22}$ (unshaded-unshaded) 
$=$ $t$ and $t^{12}$ (shaded-unshaded)$=0$. 

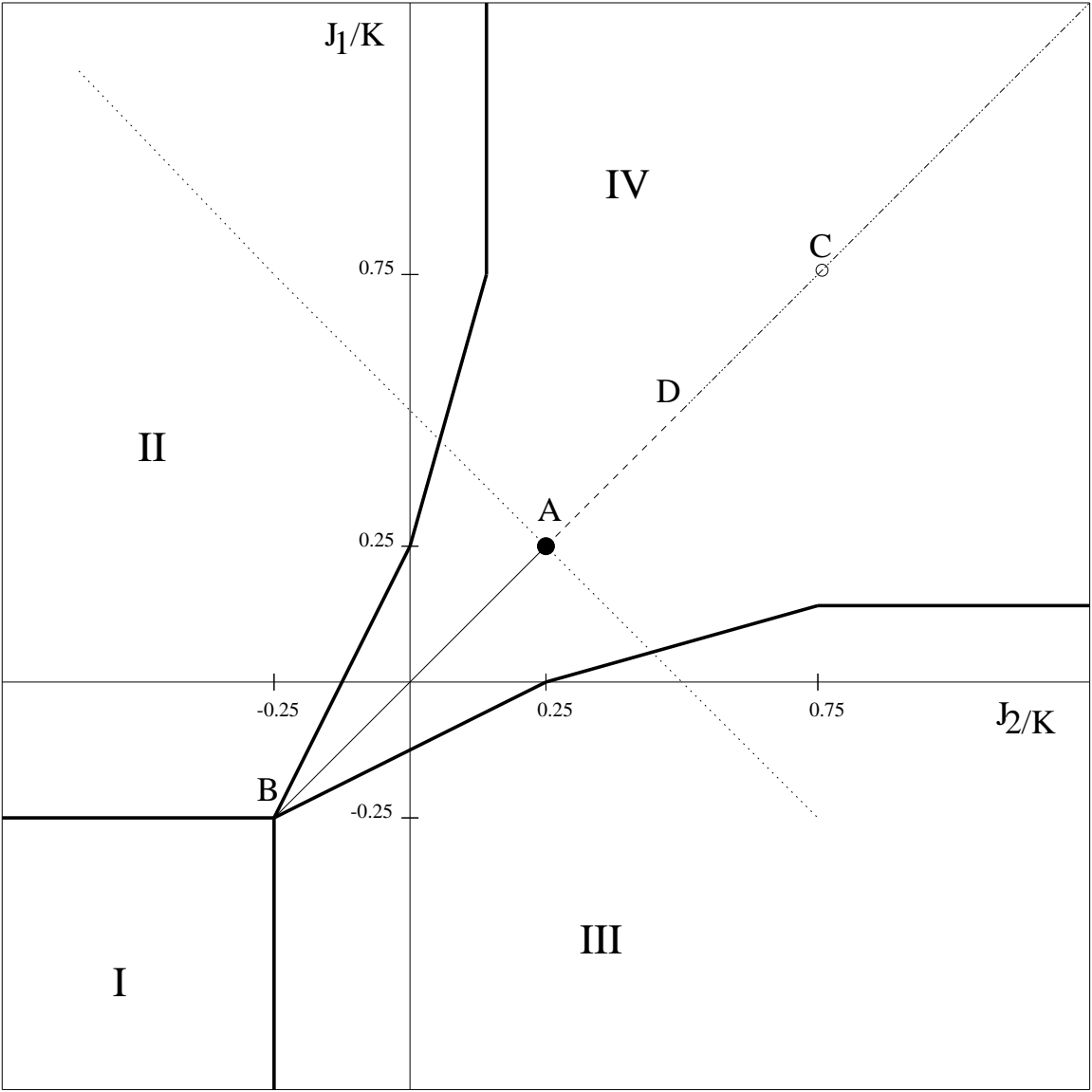
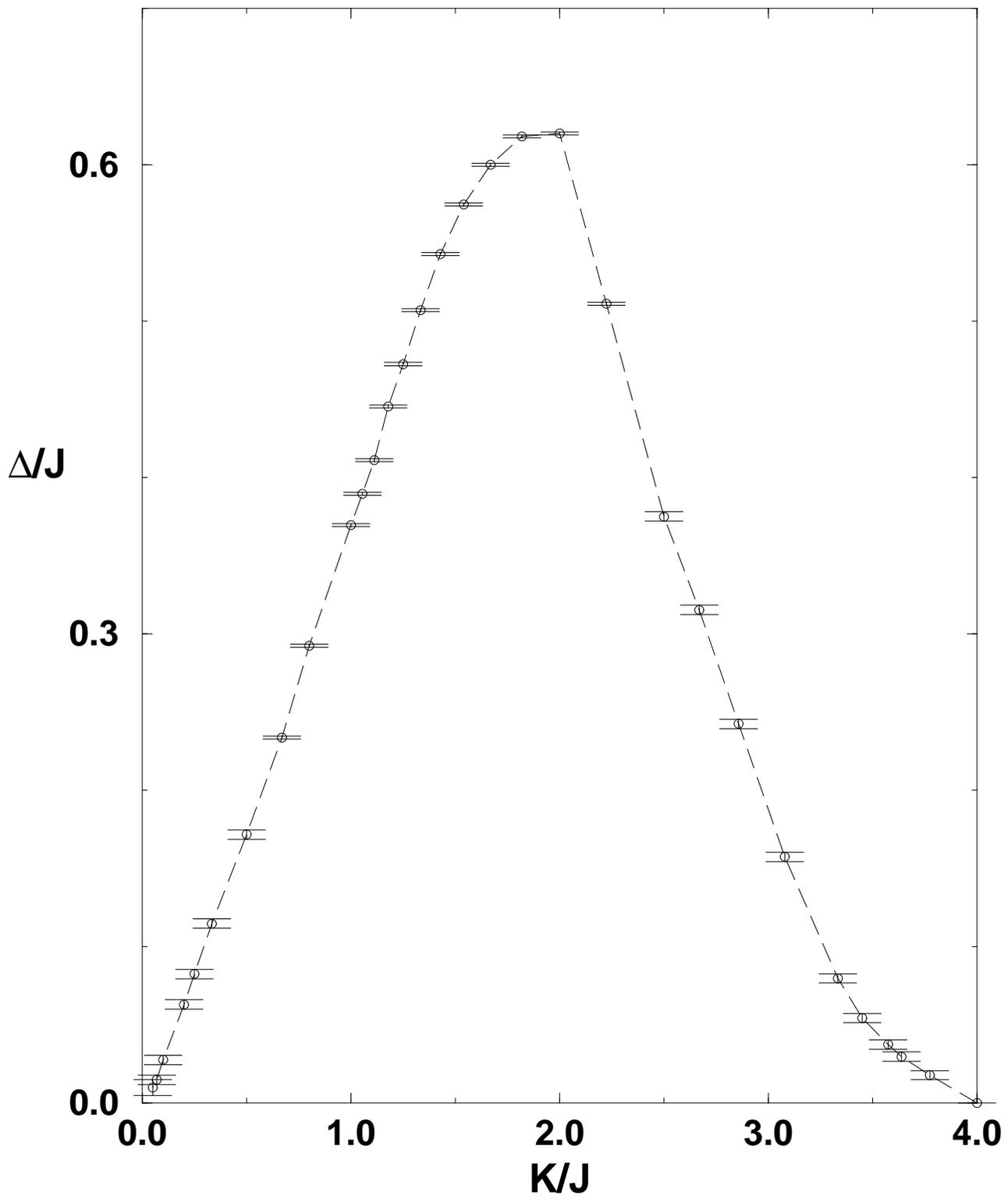
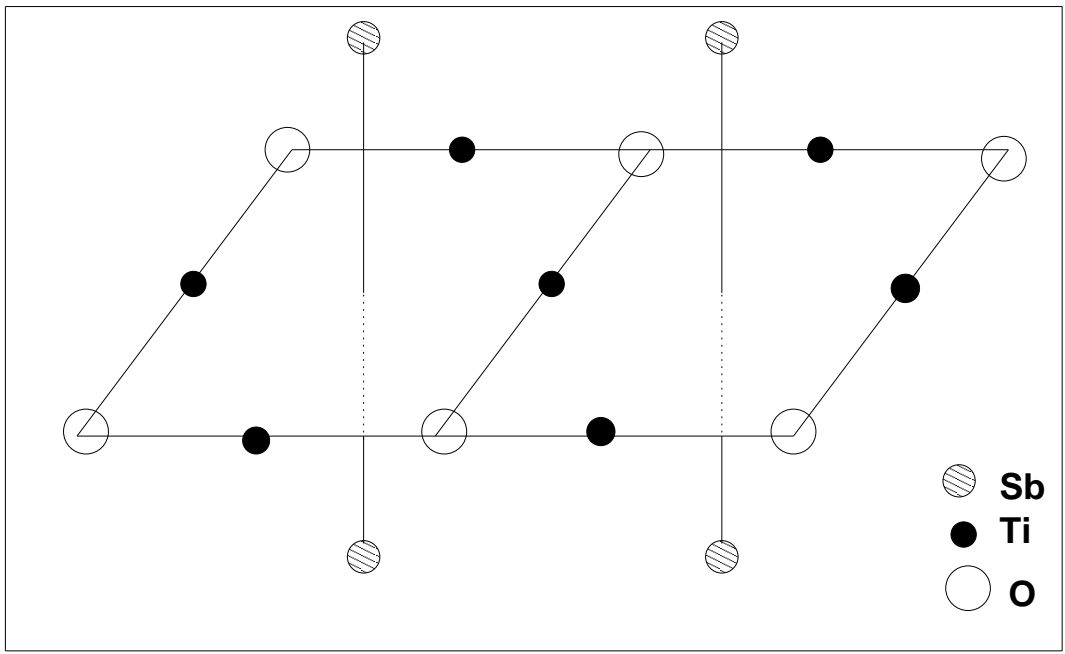
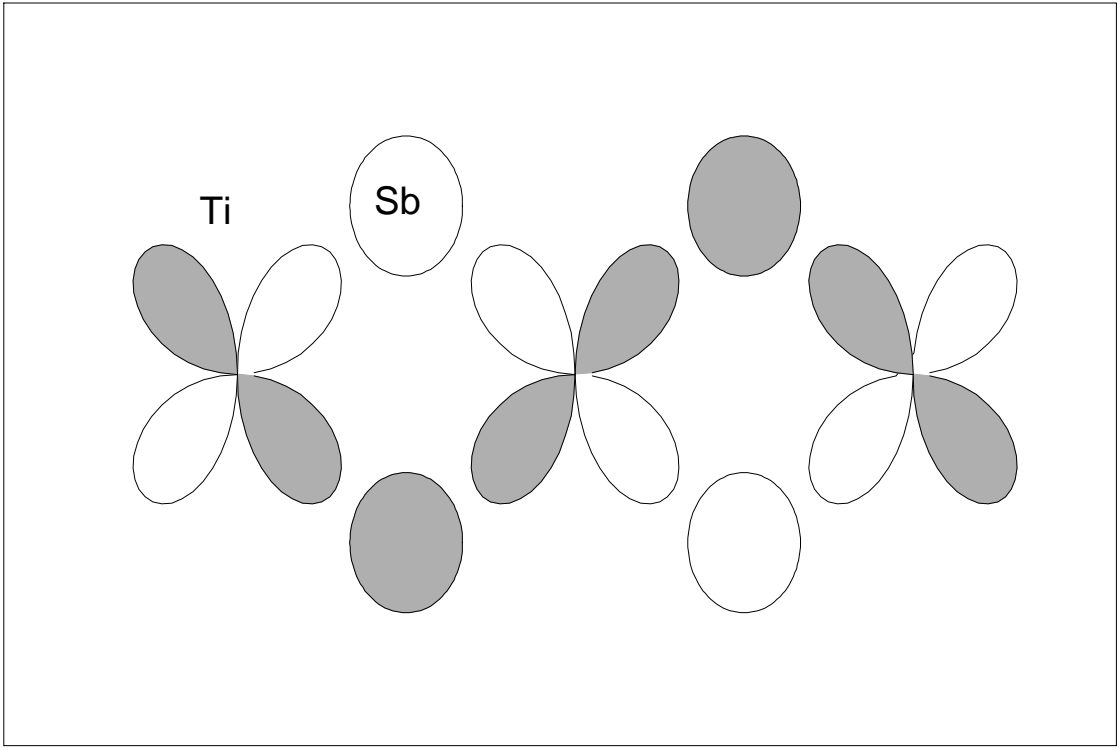
\end{document}